\documentclass[twocolumn,prb,showpacs,floatfix]{revtex4}  
\usepackage{amssymb,latexsym,stmaryrd,multirow}
\usepackage{verbatim,graphicx,subfigure,color}

\DeclareMathSizes{11}{11}{7}{5} 

\newcommand{\dirsep}{/}

\begin{document}

\DeclareGraphicsExtensions{.pdf,.ps,.eps}

\title[Ga self-interstitial relaxation in GaAs]
      {Gallium self-interstitial relaxation in Gallium Arsenide:\ an \emph{ab initio} characterization}
\author{Marc-Andr\'e Malouin} \email{marc.andre.malouin@umontreal.ca}
\author{Fedwa El-Mellouhi}\email{f.el.mellouhi@umontreal.ca} 
\author{Normand Mousseau} \email{Normand.Mousseau@umontreal.ca}
\affiliation{D\'epartement de physique and Regroupement qu\'eb\'ecois sur les mat\'eriaux de pointe,
  Universit\'e de Montr\'eal, C.P. 6128, succursale Centre-ville, Montr\'eal (Qu\'ebec) H3C 3J7, Canada}

\date{\today}

\begin{abstract}

Ga interstitials in GaAs ($I_{Ga}$) are studied using the local-orbital \emph{\mbox{ab-initio}}
code SIESTA in a supercell of \mbox{216+1} atoms. Starting from eight different initial
configurations, we find five metastable structures: the two tetrahedral sites in
addition to the \mbox{110-split$\mathrm{_{[Ga-As]}}$}, \mbox{111-split$\mathrm{_{[Ga-As]}}$}, and \mbox{100-split$\mathrm{_{[Ga-Ga]}}$}. Studying the competition between various configuration and charges of $I_{Ga}$ at T~=~0~K, we find  that predominant gallium interstitials in GaAs are charged $+1$,   neutral  or at most $-1$ depending on doping conditions and prefer to occupy the tetrahedral configuration  where it is surrounded by Ga atoms. Our  results are in excellent agreement with recent experimental results concerning the dominant charge of $I_{Ga}$, underlining the importance of finite size effects in the calculation of defects.
\end{abstract}

 
\maketitle


\section{Introduction} \label{S:int}

Gallium self-interstitials are believed to play a significant role for dopant diffusion in GaAs.
The in-diffusion of an acceptor dopant $A^+_{I}$ (at an interstitial position) occurs via a kick-out mechanism that transforms it to a substitutional atom ($A^-_{Ga}$) and a gallium interstitial ($I^k_{Ga}$)
plus the emission of a number of holes (equation from cited Ref.~\onlinecite{Bra05}): 
\begin{equation} 
A^+_I\rightarrow A^-_{Ga} + I^k_{Ga} +(2-k)h 
\end{equation} 
where k denotes the charge state of $I^k_{Ga}$ involved in the reaction.

Early calculations for Ga self-interstitials in GaAs~\cite{Zan91, Jan88_2} led
experimental groups to propose contradicting conclusions regarding the charge
state of active self-interstitials in GaAs. The suggested states varied from
neutral~\cite{Dea89}, to $+1$~\cite{Mos03, Hu95_1, Hu95_2}, $+2$~\cite{Yu91,
Uem92,Zuc89}, or both $+2$ and $+3$~\cite{Win71,Bos95}. Recently, however, Bracht
\textit{et al.}~\cite{Bra05} found that fits of recent as well as earlier
diffusivity profiles are more accurate for dominant neutral and $+1$ charge
states. This analysis of published data is confirmed by the observed
compatibilities between the hole concentration measurements and dopant (Zn)
concentrations~\cite{Bra05}.

These experimental results demonstrate the need for a set of more detailed and
accurate quantum mechanical calculations regarding the dominant charge state and
geometry of $I_{Ga}$ in GaAs. Most recent papers only treat a subgroup of the
charge states and interstitial positions~\cite{Zol03, Vol05}, however, and we
still lack a complete description of the competition between different Ga
self-interstitials in GaAs. This paper intends to fill this gap by providing a
unified analysis of all charge states from $q=-3$ to $q=+3$ for a wide range of
the Ga self-interstitial configurations at T~=~0~K, in order to identify the dominant defects
but also characterize others that could play a role in the diffusion of $I_{Ga}$
or after ion beam implantation, for example.

This paper is organized as follows. Section~\ref{S:met} explains the methodology
used for defect calculation. Next, we describe in Section~\ref{S:cfg} the
Gallium interstitial configurations used as starting points for this work.
Section~\ref{S:res} is devoted to study the stability of the chosen gallium
interstitials after full relaxation of both the neutral and the charged states.
The most relevant Ga interstitial configurations and charge states in GaAs are
then deduced and compared with earlier results in Section~\ref{S:dis}.

\section{Methodology} \label{S:met} 

All calculations are performed using the SIESTA code~\cite{San97, Sol02} within
density functional theory (DFT) in local-density approximation (LDA). The
details of the procedure followed is discussed in our previous work~\cite{Elm05}
and we focus below on the operations and parameters specific to the Ga
self-interstitial simulations.
  
\subsection{Simulation parameters} \label{S:met:param}

Simulations are performed using a supercell with \mbox{216+1} atoms. This size is just
sufficient to prevent size effects from dominating the structure and energetics of
defects in GaAs. The wavefunctions are constructed from a double-$\zeta$
polarized basis set (DZP) and we use a $2\times2\times2$ Monkhorst-Pack mesh sampling~\cite{Elm05}. The choice of these parameters is discussed at length in our earlier work and the reader is referred to Ref.~\onlinecite{Elm05} for more details.

To test the validity of the local basis set used in this work, we
evaluate the heat of formation of bulk GaAs crystal ($\Delta H$),  defined as :
\begin{equation}
\label{E:dh}
\Delta H = \mu_{As}^{bulk} +\mu_{Ga}^{bulk} -\mu_{GaAs}^{bulk}
\end{equation}
For this, it is necessary to compute the \emph{bulk} chemical potentials,
calculated from a 32 atoms As lattice ($\mu_{As}^{bulk}$), a 64 atoms Ga lattice
($\mu_{Ga}^{bulk}$), and a 216 atoms GaAs lattice ($\mu_{GaAs}^{bulk}$)
respectively. The heat of formation represents the energy necessary to
dissociate the GaAs crystal into its Ga and As components. Table~\ref{T:mu}
compares the chemical potentials obtained using DZP with chemical potentials
derived from a plane wave basis set calculation (PW) within the \mbox{DFT-LDA} carried out by Zollo {\it et al.}~\cite{Zol04} on \mbox{64+1} atoms supercell. Our calculations provide an excellent agreement with experiment:  both for the lattice parameter and the formation enthalpy. 

\begin{table}[h]
\caption{Comparison between \mbox{DFT-LDA} calculations --- with double-$\zeta$ polarized basis set (DZP) from this work and plane waves basis set (PW) from the work of Zollo {\it et al.}~\cite{Zol04} --- and experiment for the lattice parameter (\textbf{a}), chemical potentials ($\mu$) and the resulting formation enthalpy ($\Delta H$). \emph{\mbox{Ab-initio}} calculations are performed at 0~K and experimental data at 300~K.} 
\label{T:mu}
\begin{ruledtabular}
\begin{tabular} {l | l  l  l} 
& DZP & PW~\cite{Zol04}   & Expt.~\cite{Lit03} \\
\hline
{\bf a} (\AA) & 5.60 & 5.55 & 5.65 \\
{\bf $\mu_{Ga}^{bulk}$} (eV) & -61.487 & -61.785 &\\
{\bf $\mu_{As}^{bulk}$} (eV) & -173.83 & -173.75 &\\ 
{\bf $\mu_{GaAs}^{bulk}$} (eV) & -236.05 & -236.12 &\\
\hline
{\bf $\Delta H$ } (eV) & -0.737 & -0.985 & -0.736
\end{tabular}
\end{ruledtabular}
\end{table}

\subsection{Formation energy calculations} \label{S:met:calcul}

Ga self-interstitials are placed at various sites in the supercell. Since these
positions do not necessarily correspond to a local minimum, the network is
slightly distorted and the configuration is relaxed at T~=~0~K until a total
force threshold of $0.04$~eV$/$\AA\ is reached.

The formation energy ($E_{f}$) of each self-interstitial is calculated using
\begin{equation}
\label{E:ef}
E_f = E^{'}_{f} + q(E_V +\mu_{e}) - {1\over 2}(n_{As} -n_{Ga})\Delta \mu
\end{equation}
where $E^{'}_{f}$ is the formation energy independent from doping and growing
conditions, the next term on the right-hand side depends on the doping of the sample $\mu_{e}$ (i.e. Fermi level), the charge state of the defect $q$ and the position of the valence band maximum, $E_V$; the last term is associated with the stoichiometry of the supercell containing $n_{As}$ Arsenic and $n_{Ga}$ Gallium atoms. Finally, the chemical potential difference $\Delta \mu$ is defined
as:
\begin{equation}
\label{E:du}
\Delta \mu = (\mu_{As}-\mu_{Ga})-(\mu_{As}^{bulk} -\mu_{Ga}^{bulk}),
\end{equation}

The independent formation energy can thus be calculated numerically using the
relation:
\begin{eqnarray}
\label{E:ef'}
E^{'}_{f} &=& E_{tot}(q) -{1\over 2} (n_{As} +n_{Ga})\mu_{GaAs}^{bulk}
-  \nonumber \\ & & 
{1\over 2}(n_{As} -n_{Ga})(\mu_{As}^{bulk} -\mu_{Ga}^{bulk})  
\end{eqnarray}
where $E_{tot}(q)$ correspond to the total energy of the relaxed supercell
containing the self-interstitial.

The total energy of the relaxed supercell must be corrected for the strong
perturbation produced by the net charge on the relaxed state symmetry and local
electronic properties of the supercell. We can account for the electrostatic
interaction between the charged defect and its periodic images by adding a
neutralizing \emph{jellium} background then correcting the relaxed energy
($E_{tot}(q)$). Madelung correction due to the periodic boundary conditions is
introduced following the Makov and Payne approximate procedure~\cite{Mak96}.
According to our previous work~\cite{Elm05} the monopole-monopole interactions
correction is calculated to be $0.094$~eV, $0.37$~eV, and $0.84$~eV for
charge states $\pm 1$, $\pm 2$, and $\pm 3$ respectively, while higher order
corrections were found to be negligible. Charged state formation energies, from
Section~\ref{S:res:energy} and later, were adjusted using these corrections.

Finally, the position of the Fermi level $\mu_e$ varies with doping and
temperature and depends strongly on the carrier concentration. Thus, majority
carriers (electrons or holes) can get trapped at defect levels changing the
charge state of a given defect from $q_1$ to $q_2$. The thermal ionization
energy from a charge $q_1$ to $q_2$ is by definition the value of the Fermi
level where the transition occurs:
\begin{equation}
\label{E:ioe}
E_{q_1/q_2} = \frac{E_{tot}(q_2) - E_{tot}(q_1) - (q_1 - q_2) E_V}{|q_1-q_2|}
\end{equation}
We use Eq.~\ref{E:ioe} in Section~\ref{S:res:doping} to calculate ionization energies of charged defects for metastable configurations.

\section{Initial configurations for the Ga interstitial} 
\label{S:cfg}

\begin{figure*}
\centerline{\includegraphics[width=1.0\textwidth]{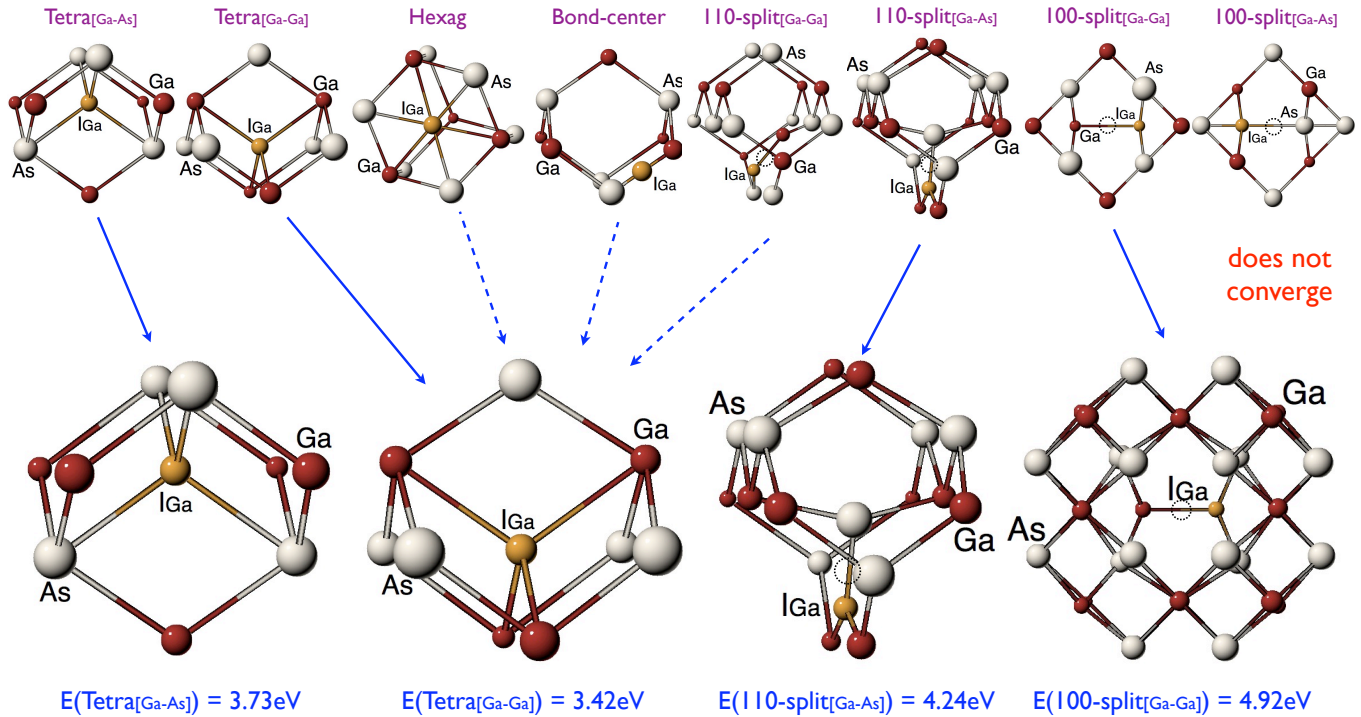}}
\caption{(Color online) Top: The eight initial configurations considered in this study for neutral self-interstitial $I_{Ga}$. The six first configurations, going from left to right, are viewed near the $<$110$>$ direction, while the remaining two are viewed along the $<$100$>$ direction. Bottom: The metastable configurations obtained after full relaxation of $I_{Ga}$. Full arrows connect the initial configuration to its metastable counterpart, while dashed arrows means that the initial configuration is unstable and converged to the pointed configuration. Gallium atoms are red while arsenics are white; the interstitial Ga atom is colored yellow. For splits, regular sites of the displaced lattice atoms are highlighted by a dotted circle.}
\label{F:config}
\end{figure*}

We first determine the metastable configurations for the Ga interstitial
(I$_{Ga}$) with different charge states. We start in each of eight positions,
relaxing the interstitial and characterizing the local energy minimum reached.
All the initial states are shown on the top row of Fig. ~\ref{F:config}. The
first starting point, from the left, tetra$\mathrm{_{[Ga-As]}}$, has the Ga
interstitial placed in a tetrahedral position with four surrounding lattice As
atoms. For the second starting configuration, tetra$\mathrm{_{[Ga-Ga]}}$, the
I$_{Ga}$ is shifted in a tetrahedral position with respect to 4 Ga atoms. In
these two initial states, the bonds between the interstitial atom and its four
tetrahedral neighbors, positioned on a perfect tetrahedron, have the same length
of $2.425\ $\AA, which is exactly the length of Ga$-$As bonds in the zincblende
structure corresponding to the lattice parameter we use $a=5.60\ $\AA\ (see
Table~\ref{T:mu}).

The third starting point is an \emph{\mbox{hexagonal}} interstitial configuration
(hexag), where I$_{Ga}$ is located at the center of the six-membered ring with
alternating chemical species at equal distances from its six nearest neighbors.
The I$_{Ga}-$Ga and I$_{Ga}-$As bond lengths are $2.322\ $\AA\ while
I$_{Ga}-$Ga$-$As angles all have the same value of $63.0^\circ$. We also
examined a \emph{\mbox{bond-center}} configuration in which I$_{Ga}$ is lying exactly
in the middle of a Ga$-$As bond at $1.213\ $\AA\ from each of them.

Finally, we look at four different interstitials from the important family of
\emph{split} geometries. Split interstitials are formed when $I_{Ga}$ pushes one
regular lattice atom (Ga or As) out of its crystalline position, forming a
dumbbell centered at a regular lattice site. The
split-interstitial type is determined by the orientation of the vector joining
the pair of atoms (see Ref.~\onlinecite{Cha92}). Gallium interstitials can form
dumbbells with As and Ga atoms following the $<$100$>$ and $<$110$>$ directions,
as shown in the top part of Fig.~\ref{F:config}, or the $<$111$>$ direction.

In more details, in the \mbox{110-split$\mathrm{_{[Ga-As]}}$} interstitial, an arsenic
atom is moved by $1.570\ $\AA\ from its regular lattice site to make room for
the intserstitial atom positioned at $0.840\ $\AA\ from the regular lattice site
and forming a dumbbell length of about $2.312\ $\AA\ along the $<$110$>$
direction. In this case, the center the dumbbell is slightly displaced from
regular lattice site (as clearly seen in  Fig.~\ref{F:s110-111}(a)).

For \mbox{110-split$\mathrm{_{[Ga-Ga]}}$}, a lattice Ga atom is moved by $1.338\ $\AA\
from its lattice site along the dumbbell axis (in another $<$110$>$ direction)
and the interstitial is placed at $0.641\ $\AA\ in the opposite direction along
this axis from the lattice site. The dumbbell length is now $1.980\ $\AA. For \mbox{100-split$\mathrm{_{[Ga-Ga]}}$} and \mbox{100-split$\mathrm{_{[Ga-As]}}$}
interstitials, the lattice atom (Ga and
As respectively) is moved along the dumbbell axis from its regular position by
$1.212\ $\AA\ while the interstitial is placed at the same distance in the
opposite direction along the same axis, forming a dumbbell of $2.425\ $\AA.

Other possible configurations including interstitial clusters might exist in real crystals. As a first step, we restricted ourselves to these simplest structures.  

\section{Results} \label{S:res}

Here we present the results of our simulations, using the techniques and
parameters described in Section~\ref{S:met}. We first discuss the stability of
the eight interstitial positions described in Section~\ref{S:cfg} in the neutral
state. Then, we focus on the influence of the charge state on the metastable
interstitial positions identified earlier. Finally, we discuss the
impact of the progressive doping of the material on the competition between
various charge states of a given interstitial in stoichiometric GaAs ($\Delta\mu
= 0$ ).

\subsection{Structural stability of neutral self-interstitials} \label{S:res:energy}

The bottom part of Fig.~\ref{F:config} shows the final geometry of the relaxed
interstitial configurations in the neutral state. The most stable interstitial
is the tetra$\mathrm{_{[Ga-Ga]}}$, which undergoes small lattice distortions
leading to its convergence into a metastable configuration with a formation energy
of approximately $3.42$~eV. The hexag, \mbox{bond-center} and \mbox{110-split$\mathrm{_{[Ga-Ga]}}$}
configurations are unstable and relax to the same tetra$\mathrm{_{[Ga-Ga]}}$ 
after undergoing considerable atomic displacement and lattice relaxation.

The second tetrahedral configuration, tetra$\mathrm{_{[Ga-As]}}$, is also
metastable, with a formation energy of about $3.73$~eV, slightly above that of
tetra$\mathrm{_{[Ga-As]}}$. This structure is only a few relaxation steps away
from the initial tetra$\mathrm{_{[Ga-As]}}$. Both tetrahedral interstitials leave the surrounding crystalline network
relatively unaffected. The tetra$\mathrm{_{[Ga-Ga]}}$ configuration is close to
the starting configuration with only a slight outward relaxation of the
surrounding Ga neighbours leading to an increase in length of both I$_{Ga}-$Ga bond by about 7.0 \%,
to $2.596\ $\AA. The volume expansion around the tetra$\mathrm{_{[Ga-Ga]}}$ goes down rapidly and
affects only the first and second neighbor shells along the tetrahedral axes.
The tetra$\mathrm{_{[Ga-As]}}$ configuration undergoes a similar expansion and
the I$_{Ga}-$As bond lengths are stretched by 5.3 \% from $2.425$ to $2.554\ $\AA.

Both the \mbox{110-split$\mathrm{_{[Ga-As]}}$} and \mbox{100-split$\mathrm{_{[Ga-Ga]}}$} are
found to be metastable. The formation energies are higher: $E_f^{'}=4.24$~eV for \mbox{110-split$\mathrm{_{[Ga-As]}}$} and $E_f^{'}=4.92$~eV for \mbox{100-split$\mathrm{_{[Ga-Ga]}}$}. In addition,
the stress imposed on the lattice is more important, and affects significantly the more distant neighbours.

The \mbox{100-split$\mathrm{_{[Ga-Ga]}}$} experiences the largest lattice deformation
around the defect among other interstitial defects as illustrated in
Figure~\ref{F:config} where surrounding lattice atoms are shown. This relaxed
configuration is a dumbbell formed by $I_{Ga}$ and the displaced Ga lattice
atom. The dumbbell is centered and symmetric with respect to the middle vacant
lattice gallium site with each Ga atom being located at $1.10\ $\AA\ apart. The
length of the dumbbell shrinks by about $9.2\%$ (from $2.42$ to $2.20\ $\AA)
bringing the two atoms closer. For their part, atoms at the first and second
shell neighbors experience an outward relaxation and are pushed away from their
original position by about $0.46$ and $0.21\ $\AA\ respectively. Thus,
considering both effects the bond length between each of the Ga atoms forming
the dumbbell and their first As lattice neighbors increases by approximately
$17.4\%$ from $1.99$ to $2.34\ $\AA\, as the distance to the second nearest
neighbors grows from $3.22$ to $3.47\ $\AA, a change of about $7.8\%$.

Finally, the 100-split$_{[Ga-As]}$ self-interstitial is highly unstable and does
not converge to any stable state. For this reason, we will not attempt any
further calculation using this configuration for the rest of this work.

In order to test the stability of the four metastable configurations found, we
further relaxed them with a more accurate force threshold of $0.002$~eV$/$\AA.
The observed change in geometry and formation energy is negligible indicating
that our results are already well converged with the former force threshold of
$0.04$~eV$/$\AA\ .

\begin{table}
\caption{Formation energies (in eV) for stable and metastable  Ga interstitial configurations in GaAs for various  charge states.}
\label{T:efc}
\begin{ruledtabular}
\begin{tabular} {l | l  l  l  l  l  l  l}
Stable & \multicolumn{7}{c}{Net system charge \emph{q}} \\
\cline{2-8}
configurations & $-3$ & $-2$ & $-1$ & \ $0$ & $+1$ & $+2$ & $+3$ \\
\hline
{tetra$_{[Ga-Ga]}$}  & 8.37 & 6.40 & 4.62 & 3.42 & 2.40 & 2.81 & 3.43 \\
{tetra$_{[Ga-As]}$} & 8.65 & 6.72 & 4.98 & 3.73 & 2.67 & 2.91 & 3.41 \\
\hline
{110-split$_{[Ga-As]}$} & \ --- & 6.42 & 4.99 & 4.24 & \ \ $\downarrow$ & \ \ $\downarrow$ & \ \ $\downarrow$\\
{111-split$_{[Ga-As]}$} & \ --- &          &          &          & 3.33 & 3.74 & 4.35\\
\hline
{100-split$_{[Ga-Ga]}$} & \ --- & 7.75 & 6.16 & 4.92 & 3.33 & 4.15 & 4.62 \\

\end{tabular}
\end{ruledtabular}
\end{table}	

\subsection{Structural stability of charged self-interstitials} \label{S:res:charged}

Having characterized the stability of the set of neutral initial self-interstitial configuration,
we now turn to charge states. All charge configurations are also started from the eight ideal
unrelaxed configurations except for the \mbox{110-split$\mathrm{_{[Ga-As]}}$}, which could
not relax in the allowed time from the ideal position and which was started in the neutral
relaxed configuration instead.

\begin{figure}
       \centerline{\includegraphics[width=10cm]{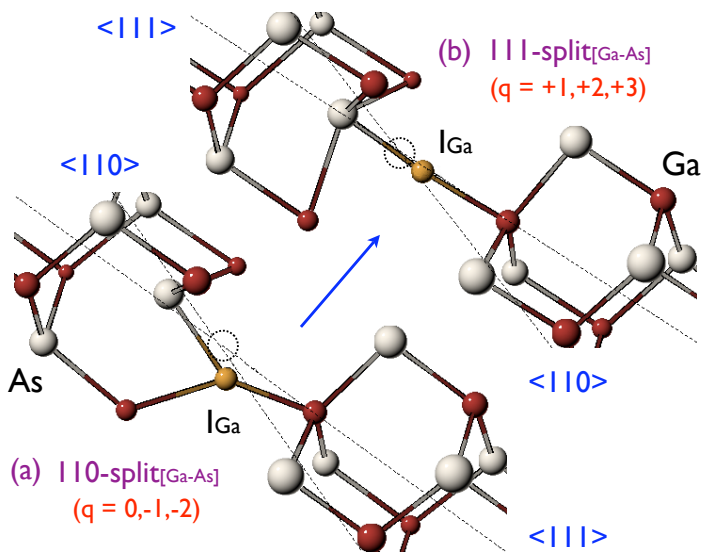}}
   	\caption{(Color online) The two distinct 110 (a) and 111 (b) splits seen from a $110$  view. Dash lines correspond to $<$111$>$ and $<$110$>$ crystalline axis (indicated in blue) and dotted circles refer  to the regular crystalline position of the displaced arsenic  atom. Gallium atoms are red while arsenics are white; the interstitial Ga atom is colored yellow. Some lattice atoms have been removed for clarity.}
	\label{F:s110-111}
\end{figure}

For $q=+1$ charged interstitials, the relaxation follows the same scenario as
for the neutral defect for all first seven configurations of Fig.~\ref{F:config}
(top): the overall stability order is kept unchanged and the
tetra$\mathrm{_{[Ga-Ga]}^{+1}}$ configuration is still the most stable defect.
For their part, the unstable interstitial states relaxed into the same
metastable configuration as in the neutral case with one exception, however: the
\mbox{110-split$\mathrm{_{[Ga-As]}}$} now relaxes into a \mbox{111-split$\mathrm{_{[Ga-As]}}$}
(Fig.~\ref{F:s110-111}) but keeps the same stability order with respect to the
other metastable defects.

Interestingly, the resulting formation energies calculated for $+1$ charged
defects are significantly lower than in the neutral charge state in all cases:
tetra$\mathrm{_{[Ga-Ga]}^{+1}}$ has a formation energy of $E^{'}_{f}=2.40$~eV,
followed by tetra$\mathrm{_{[Ga-As]}^{+1}}$ with $E^{'}_{f}=2.63$~eV, then
111-split$\mathrm{_{[Ga-As]}^{+1}}$ with $E^{'}_{f}=3.33$~eV and finally
100-split$\mathrm{_{[Ga-Ga]}^{+1}}$ with also $E^{'}_{f}=3.33$~eV.

The removal of an electron for $I_{Ga}$ in GaAs stabilizes all defects uniformly
with respect to their respective neutral state while the stability order of each
interstitial configuration with regard to each other remains about the same. To
fully characterize this effect, we have further relaxed the most stable
interstitial geometries --- tetra$\mathrm{_{[Ga-Ga]}}$,
tetra$\mathrm{_{[Ga-As]}}$, \mbox{100-split$\mathrm{_{[Ga-Ga]}}$}, and 110 or 111
splits$\mathrm{_{[Ga-As]}}$ --- for the $q=-1$, $q=\pm2$ and $\pm3$ charges,
supposing that the unstable interstitial configurations do not stabilize in
these highly charged systems.

Table~\ref{T:efc} shows the formation energies of the five metastable relaxed
configurations in increasing order of formation energy, for seven charge states
($q=\pm3, \pm2, \pm, \text{and}\ 0$) with the associated monopole
correction applied, as explained in Section~\ref{S:met:calcul}. The missing
numbers for charge $q=-3$ correspond to configurations that did not achieve
convergence even after long simulations. Additionally, starting from the the neutral \mbox{110-split$\mathrm{_{[Ga-As]}}$}
configuration, relaxations towards positive charges all induce a change in the orientation of the dumbbell leading to the
nearly same \mbox{111-split$\mathrm{_{[Ga-As]}}$} (as indicated by downside arrows in Table~\ref{T:efc}) while relaxations of negatively charged \mbox{110-split$\mathrm{_{[Ga-As]}}$} preserve the initial $<$110$>$ orientation.

Figure~\ref{F:s110-111}  illustrates the shift of orientation from \mbox{110-split$\mathrm{_{[Ga-As]}}$} for neutral and negative charges to \mbox{111-split$\mathrm{_{[Ga-As]}}$} for positive charges. The length of the  dumbbell  is around $2.24\ $\AA\ for 111-split$\mathrm{_{[Ga-As]}}^{+1,+2,+3}$ and about  $2.33\ $\AA\ for 110-split$\mathrm{_{[Ga-As]}}^{0,-1,-2}$. 
 
From Table ~\ref{T:efc}, we observe that the lowest formation energy for all the
four interstitial configurations is associated with the $+1$ charged state. For
a given symmetry, we see that the formation energy monotonically decreases with
increasing charge, from $-3$ to $+1$ before going up for more positive charge
states. In all cases, the formation energies for positive charges are lower than
for the negative ones. 

Looking at particular charged interstitials, we observe also that the order of
stability between different interstitial configurations that we observe in
neutral charge state is conserved under the variation of the net charge of the
system (except for $-2$ charged state where \mbox{110-split$\mathrm{_{[Ga-As]}}$}
formation energy is below that of tetra$\mathrm{_{[Ga-As]}}$). Moreover, apart
from higher charge states, there is a somewhat constant difference of $0.30$~eV
between the two different tetrahedral configurations for the same charge states
from $-3$ to $+1$ where the formation energy reaches its minimum value.

We note also that a degeneracy appears between pairs of interstitials for two
charge states: at $+1$, both 111 and 100 splits have about the same formation
energy while at $-1$, the same is true for tetra$\mathrm{_{[Ga-As]}}$ and
\mbox{110-split$\mathrm{_{[Ga-As]}}$}. This suggests that the 110 and 111
split$\mathrm{_{[Ga-As]}}$ configurations play a key role in GaAs crystals,
serving as transitional configurations when passing from
tetra$\mathrm{_{[Ga-As]}}$ at charge $-1$ to \mbox{100-split$\mathrm{_{[Ga-Ga]}}$} at
charge $+1$, successively losing two electrons, one at a time (the 111-split
being an intermediate step from neutral to $+1$ charged state). As a result,
this specific transition process could be an important diffusion path for
impurities in GaAs crystals.

Although, for charge $+3$, the formation energies of both tetrahedral configurations are also near, it could be a finite size effect associated with the stress induced by the high positive net charge, introducing a  bias in our calculation.

\subsection{Competition between $I_{Ga}$ charge states under doping conditions} \label{S:res:doping}

\begin{figure}
        \centerline{\includegraphics[width=10cm]{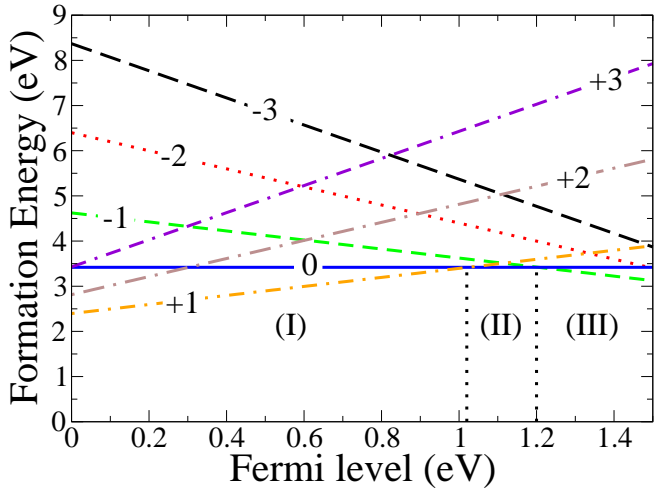}}
   	\caption{(Color online) Formation energies as function of Fermi level for various charge states of the tetra$\mathrm{_{[Ga-Ga]}}$ configuration at 0~K.  Dotted lines point at the ionization level locations  delimiting the three stability domains labeled by (I), (II), and (III).}
 	\label{F:g:t}
\end{figure}

\begin{figure}
        \centerline{\includegraphics[width=10cm]{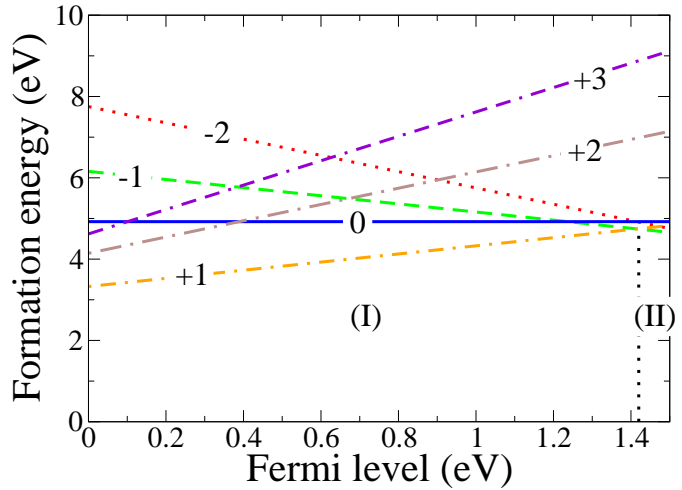}}
    	    \caption{(Color online) Formation energies as function of Fermi levels for various charge states of the \mbox{100-split$\mathrm{_{[Ga-Ga]}}$} configuration at 0~K.}
	\label{F:g:s100a}
\end{figure}

\begin{figure}
          \centerline{\includegraphics[width=10cm]{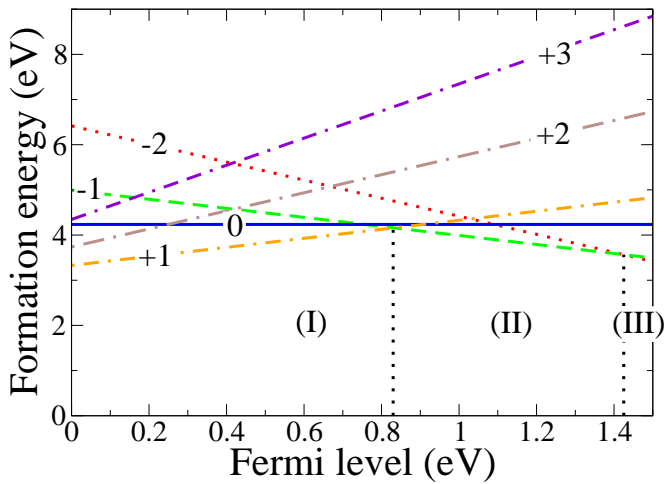}}
          	\caption{(Color online) Formation energies as function of Fermi levels for various charge states of the 110 ($q=-2,-1,0$) and 111 ($q=+1,+2,+3$) split$\mathrm{_{[Ga-As]}}$ configuration at 0~K.}
	\label{F:g:s110}
\end{figure}

We now look at the effect of doping by varying the Fermi level with the help of
the parameter $\mu_e$ in equation~\ref{E:ef}. These effects are best seen by
comparing data for multiple charge states and we concentrate on the
configurations of Table~\ref{T:efc}. Because of the similarities in the
stability diagrams of tetra$\mathrm{_{[Ga-Ga]}}$ and tetra$\mathrm{_{[Ga-As]}}$
only the first one is shown in Figure~\ref{F:g:t}. The diagram for
\mbox{100-split$\mathrm{_{[Ga-Ga]}}$} and \mbox{110-split$\mathrm{_{[Ga-As]}}$} are shown in
Figures~\ref{F:g:s100a} and~\ref{F:g:s110} respectively. In all figures, the
Fermi level is set by reference to the valence band maximum.

Because DFT calculations are known to underestimate the band gap (the present
calculation gives a band gap of $0.82$~eV compared to $1.52$~eV at T~=~0~K
reported by experiment), it is common to vary the Fermi level in the window of
the experimental band gap to obtain the full picture~\cite{Wal04}.
Note, that even if our calculated ionization levels have a conduction band
character, they have not been corrected for that because the efficiency of the  several band gap correction methods is system dependent and testing them is out of the scope of the present work~\cite{Cas06}.

Fig.~\ref{F:g:t} shows the formation energy as a function of the doping level for the tetrahedral insterstitial.  Dotted lines point at the location of ionization level, identified by intersecting formation energy lines. Three stability domains, labeled by  (I), (II), and (III), are found for  tetra$\mathrm{_{[Ga-Ga]}}$ corresponding to  successive dominant charge states $+1$, $0$, and $-1$ respectively. The same stable states occur for tetra$_{[Ga-As]}$, with almost the same ionization energies. Table~\ref{T:iol} summarizes these ionization energies, calculated from equation~\ref{E:ioe}.

The \mbox{100-split$\mathrm{_{[Ga-Ga]}}$} exhibits a different behavior as the line
of charge state $q=-1$ crosses the line from the state $q=+1$ \emph{before}
the horizontal neutral $q=0$ state line (about $0.18$~eV below) allowing a
direct transition between $q=-1$ and $+1$, in a so-called \emph{\mbox{negative-U}}
effect. Only two stability domains labeled by (I) and (II) are found, for
charge $+1$ and $-1$ respectively (see Fig.~\ref{F:g:s100a}). This effect
might not exist in real systems since region II occurs at the edge of the
conduction band. Any fluctuation or error might therefore screen region II.
The reverse could be true for the tetrahedral interstitials
(Fig.~\ref{F:g:t}). In this figure, region II is very narrow, set in the
middle of the gap, and a fluctuation could remove it altogether, this time
inducing a \mbox{negative-U} effect.

We note also that the transition domain of the 100-split occurs just below the
minimum of the conduction band, meaning that it is only accessible in extreme
doping condition. For most purposes, the domain of charge $+1$ will be the
only one that matters.

Another \mbox{negative-U} effect causes the transition from
111-split$\mathrm{_{[Ga-As]}}^{+1}$ to 110-split$\mathrm{_{[Ga-As]}^{-1}}$
(Fig.~\ref{F:g:s110}). Once again, the transition occurs very near to the
neutral charge state line, only $0.07$~eV below. Here, however, the transition
is located at midgap and should therefore play a more important role.
Moreover, as this ($+1$/$-1$) \mbox{negative-U} transition manifests itself by a
change in the orientation of the dumbbell from 111 to 110, it should therefore
be relatively insensitive to the various limitations of the current simulation
and other possible thermal fluctuations.

Similarly to the 100-split, we find also a transition (here, $-1$/$-2$) for
\mbox{110-split$\mathrm{_{[Ga-As]}}$} near the conduction band minimum. Since this
transition level has itself a conduction band character its position might be
affected by the DFT-bandgap underestimation and must be treated with care.
Although we did not manage to converge the
110-split$\mathrm{_{[Ga-As]}}^{-3}$, it is interesting to note that if the
trend for the negative charges seen in Fig.~\ref{F:g:s110} holds, we should
see a \mbox{negative-U} transition between charge $-1$ and $-3$ before the $-1$/$-2$
transition for this charge.

\begin{table}
\caption {Ionization energies of   metastable $I_{Ga}$ configurations in GaAs (see equation~\ref{E:ioe}). \mbox{\mbox{Negative-U}} transition ($+1$/$-1$) for split$\mathrm{_{[Ga-As]}}$ changes the dumbbell orientation from 111  to 110.}
\label{T:iol}
\begin{ruledtabular}
\begin{tabular} {l c c c c}
& \multicolumn{4}{c}{Ionization levels (eV)} \\
\cline{2-5}
& & & \emph{\mbox{Negative-U}} & \\
Configurations & $+1$/$0$ & $0$/$-1$ & $+1$/$-1$ & $-1$/$-2$\\
\hline
tetra$\mathrm{_{[Ga-Ga]}}$ &1.02 & 1.20 & &\\
tetra$\mathrm{_{[Ga-As]}}$ & 1.06 & 1.25 & &\\
\hline
\mbox{100-split$\mathrm{_{[Ga-Ga]}}$} & & & 1.42 &\\
\hline
110 \& 111 splits$\mathrm{_{[Ga-As]}}$ & & & 0.83 & 1.42 \\
\end{tabular}
\end{ruledtabular}
\end{table}

%

\section{Discussion} \label{S:dis}

Here we discuss and compare our results with previous \emph{ab initio} and
tight-binding calculations~\cite{Cha92,Zol03, Zol04,Vol05}, as well as with
recent experimental data from Bracht~\cite{Bra05}. While it was not clearly
indicated in most of the theoretical works, whether the \emph{tetrahedral}
interstitial label meant tetra$\mathrm{_{[Ga-As]}}$ or
tetra$\mathrm{_{[Ga-Ga]}}$, we will presume that it refers to the former.

Chadi~\cite{Cha92} was first to report self-interstitial configurations and
energetics using self-consistent pseudo-potential relaxations on GaAs
supercells with \mbox{32+1} atoms. Almost the same set of starting configurations as
in the present work was studied under different charging, but the resulting
stability order was completely different from our. Indeed, Chadi found the
\mbox{bond-center} (\emph{twofold}) configurations to be the most stable configuration
for $q=+1$ and the \mbox{110-split$\mathrm{_{[Ga-As]}}$} for $q=0,-1$. The
tetra$\mathrm{_{[Ga-As]}}$ was found to have the lowest formation energy only
under $+2$ charging. The difference between this work and ours is mostly due
to the strong finite size effects present in a \mbox{32+1} unit cell. 

For their part, Zollo and Nieminen~\cite{Zol03} studied the full set of interstitial positions
--- except for $tetra_{[Ga-Ga]}$--- in a \mbox{64+1} atomic supercell and for the
neutral state only. They found that the hexagonal, \mbox{bond-center}, and 100-splits
are unstable, converging to the tetrahedral interstitial position. They also
identified the tetra$\mathrm{_{[Ga-As]}}$, the \mbox{110-split$\mathrm{_{[Ga-Ga]}}$},
and the \mbox{110-split$\mathrm{_{[Ga-As]}}$} to be metastable, with increasing
formation energy. Recently, Volpe {\it et al.}~\cite{Vol05} used a large
supercell of \mbox{216+1} atoms with tight binding method to study $I_{Ga}$ in GaAs,
again treating exclusively the neutral charge state and computing formation
energies relative to the tetrahedral interstitial formation energy only. The
metastable structures identified at the neutral state were classified in
increasing order of formation energy: the tetra$\mathrm{_{[Ga-As]}}$, then the
\mbox{\mbox{110-split$\mathrm{_{[Ga-Ga]}}$}}, the \mbox{110-split$\mathrm{_{[Ga-As]}}$}, and the
\mbox{100-split$\mathrm{_{[Ga-Ga]}}$}. As was shown in Tab.~\ref{T:efc}, our results
show that \mbox{100-split$\mathrm{_{[Ga-Ga]}}$} is indeed metastable as found by
Volpe {\it et al.}. They disagree with both Zollo and Nieminen and Volpe {\it
et al.} with respect to the \mbox{110-split$\mathrm{_{[Ga-Ga]}}$} interstitial,
however, which becomes unstable and prefers to relax to
tetra$\mathrm{_{[Ga-As]}}$ according to our 216 DFT calculations.

The difference between these calculations and the one presented here are
caused by (1) size effect associated self-interactions of the defects in unit
cells that are too small and (2) the accuracy of the potential (DFT versus
tight-binding), particularly when important structural changes are taking
place, as is the case for split interstitials, for example.

Focusing on the dominant charge state of $I_{Ga}$ in GaAs, the results of
Section~\ref{S:res} show that higher charge states ($q=\pm2, \pm3$) are not
relevant and should contribute negligibly to the total experimental
self-diffusion profiles. This agrees well with recent experimental results
from Bracht~\cite{Bra05} that identified $I^0_{Ga}$ and $I_{Ga}^{+1}$ as
important species for diffusion process in GaAs crystals doped with Zn (but
with lower contribution than vacancies). Our results provide a strong support
for the picture proposed by Bracht, disproving earlier models which generally
predict diffusion process controlled mainly by $+2$ and $+3$ interstitials. In
particular, our calculations show that $+2$ and $+3$ charge states exhibit
higher formation energies than the $+1$ charge defect, contrary to what was
found by Zhang and Northurp~\cite{Zan91}. Using a \mbox{32+1} atoms supercell within
\mbox{DFT-LDA}, these authors identified the dominant native defect to be the
tetra$\mathrm{_{[Ga-As]}^{+3}}$ in Ga-rich condition under p-type doping.
Similarly, more accurate \emph{\mbox{ab-initio}} calculation of Zollo {\it et
al.}~\cite{Zol04} find \mbox{negative-U} effects for tetra$\mathrm{_{[Ga-As]}}$ with
($+3$/$+1$) and ($+1$/$-1$) ionization levels located at $0.29$ and $1.23$~eV
above the VBM respectively, still giving an important role to the triply
positive state which we do not see.

Again, size effects can explain many of these divergences. For example, finite
size effects have been reported recently by Schick {\it et al.}~\cite{Sch02}
for As 110-split interstitials in GaAs. While a 65 supercell calculation with
a $2\times2\times2$ k-points mesh predict the stability of $+2$ charge state
starting for the VBM. This charge state disappears completely from the diagram
as soon as a supercell as large as 217 is used with different k-point meshes
leaving the $+1$ charge state as the most stable near VBM.

Qualitatively, the formation energies for tetra$\mathrm{_{[Ga-Ga]}}$ in
stoichiometric GaAs ($\Delta\mu=0$) we compute depend on doping conditions:
$E_f({I_{Ga}^{0}})$~=~$3.42$~eV, $E_f({I_{Ga}^{+1}})$~=~$2.4$ to $3.92$~eV,
$E_f({I_{Ga}^{+2}})$~=~$2.81$ to $4.33$~eV, and for
$E_f({I_{Ga}^{+3}})$~=~$3.43$ to $4.95$~eV. All these values remain in the
window of allowed values compared to the activation enthalpy ($H_a$) obtained
after fitting the experimental profiles. $H_a$ for $I_{Ga}^{0, +1}$-mediated
Ga diffusion in GaAs reported by Bracht~\cite{Bra05} was $5.45\pm0.12$ and
$5.80\pm0.32$~eV for neutral and +1 charge states respectively. Since we do
not know the migration enthalpies of $I_{Ga}$ in GaAs with respect to the
charge state, it is not possible at this point to push further and identify
the charge state responsible for the Zn diffusion profiles. Only a detailed
study of the migration mechanisms of $I_{Ga}$ in GaAs similar to the one
performed recently for $V_{Ga}$ in GaAs as function of the charge
state~\cite{Elm06} can give the answer to this question.

\section{Conclusions} \label{S:sum} 

In this work, we have studied the stability of Ga self-interstitial for
multiple charge states ($q=0, \pm1, \pm2, \text{and}\pm 3$) within \mbox{DFT-LDA}
using the local-orbital basis set program SIESTA at T~=~0~K. Out of the eight
initial configurations tested, five were found to be metastable after full
relaxation. As a general rule, the most stable configuration is found to be
tetra$\mathrm{_{[Ga-Ga]}}$ for all charge states, in addition positively
charged interstitials are more stable than negative ones for all tested
charges and configurations.

After studying the competition between various configuration and charges of
$I_{Ga}$, we conclude that predominant gallium interstitials in GaAs are
charged $+1$, neutral or at most $-1$ depending on doping conditions. This
agrees well with the recent conclusions driven by Bracht {\it et al.} which
states that fits of recent as well as earlier diffusivity profiles in Zn doped
GaAs are more accurate if the role $I_{Ga}^{+1,0}$ is considered. At low
temperatures, when the formation energy dominates, $I_{Ga}$ prefers to occupy
the tetrahedral interstitial configuration being surrounded by gallium atoms
(tetra$\mathrm{_{[Ga-Ga]}}$). The competition between
tetra$\mathrm{_{[Ga-Ga]}}$ and the other metastable configuration increases as
we approach the experimental processing temperatures (above 1000~K) but not
sufficiently to invert the order of stability. For example under P-type doping
at 1000~K, the tetra$\mathrm{_{[Ga-Ga]}}^{+1}$ still has a probability of
occurrence about 100 times larger than $I_{Ga}^{+1}$ and $10^{5}$ times larger
than the 110 and 100-split interstitials.

The comparison of our results with previous works show also that
size of the simulation supercell can affect significantly the stability and
formation energy of $I_{Ga}$, and both tight-binding and \emph{\mbox{ab-initio}}
calculations become more reliable with increasing cell size. The size of the
supercell affects the charge state of the dominant defect and modifies also
the metastability of other defects such as the split interstitials. In this
case, the use of a \mbox{216+1} atoms supercell allows us to observe for the first
time a change in the orientation of the split from $<$110$>$ to $<$111$>$
after the removal of one electron from the neutral configuration.

In spite of the excellent agreement with recent experimental results, further calculations,
including the entropic contributions and the diffusion pathways, are still necessary to obtain
a complete picture of the role of $I_{Ga}$ and reveal the possible importance of other
charge states and configurations.


\section{Acknowledgments} \label{S:ack}

We would like to thank Dr. Giuseppe Zollo for sharing his results on bulk As,
Ga, and GaAs chemical potentials. Dr. Harmurt Bracht for fruitful discussion.
This work is funded in part by NSERC (Canada), FQRNT (Qu\'{e}bec) and the
Canada Research Chair Foundation. All the simulations were run with the
computer network support of R\'{e}seau qu\'{e}becois de calcul et de haute
performance (RQCHP) which is greatly appreciated.

\bibliography{references_stage.bib}
\end{document}